\documentclass[reprint,amsmath,amssymb,aps]{revtex4-2}

\usepackage{graphicx}
\usepackage{dcolumn}
\usepackage{bm}
\usepackage{hyperref}
\usepackage{graphicx,amssymb,amsmath,amsfonts}
\usepackage{verbatim}
\usepackage{xcolor}

\def\be{\begin{equation}}
\def\ee{\end{equation}}

\newcommand{\alp}{{\alpha^{\prime}}}

\newcommand{\pmat}{\begin{pmatrix}}
\newcommand{\fpmat}{\end{pmatrix}}
\newcommand{\eq}{\begin{equation}}
\newcommand{\feq}{\end{equation}}
\newcommand{\cas}{\begin{cases}}
\newcommand{\fcas}{\end{cases}}

\newcommand{\eqarray}{\begin{eqnarray}}
\newcommand{\feqarray}{\end{eqnarray}}

\begin{document}

    
\title{Multi-dimensional chaos II:\\ String scattering amplitudes, curve repulsion, and RMT}

\author{Massimo Bianchi}
 \email{massimo.bianchi@roma2.infn.it}
 \affiliation{Dipartimento di Fisica, Università di Roma Tor Vergata,\\
Via della Ricerca Scientifica 1, 00133, Roma, Italy
}
\affiliation{INFN sezione di Roma Tor Vergata \\
Via della Ricerca Scientifica 1, 00133 Roma, Italy
}
 
\author{Maurizio Firrotta}
 \email{mfirrotta@fudan.edu.cn} 
\affiliation{Department of Physics and Center for Field Theory and Particle Physics, \\Fudan University, Shanghai 200438, China
}

\author{Jacob Sonnenschein}
\email{cobi@tauex.tau.ac.il}
\affiliation{The Raymond and Beverly Sackler School of Physics and Astronomy, \\
Tel Aviv University, Ramat Aviv 69978, Tel Aviv, Israel}

\author{Dorin Weissman}
\email{dorin.weissman@na.infn.it}
\affiliation{INFN Sezione di Napoli, \\Monte S. Angelo, Via Cintia, 80126 Naples, Italy}

\date{\today}

\begin{abstract}
 Multi-dimensional chaos refers to processes described by erratic functions of several dynamical variables. In this letter we analyze the string scattering amplitudes of highly-excited states and ground states. We show that the amplitudes, which depend on a scattering angle and a polarization angle, are characterized by two sets of non-intersecting curves associated with the vanishing of the derivatives with respect to the angles. We introduce the notion of the ``area eigenvalue'' $A_n$ associated with the $n$-th curve. We compute the spacings $\delta_{n}= A_{n+1}-A_n$  and their ratios $r_{n}=\frac{\delta_{n+1}}{\delta_n}$. We show that
  the distributions of the spacing ratios take the form of the RMT Gaussian $\beta$-ensembles. The curves associated with the scattering angle tend to converge to the Gaussian Orthogonal Ensemble value of $\beta=1$ and those related to the polarization angle to the Gaussian Unitary Ensemble $\beta=2$. We also compute the ``areas form factor" associated with the areas and discover the regions of decline, ramp and plateau which characterize chaotic processes. The slope of the ramp seems to agree with the $\beta$ values extracted from the distribution of the spacing ratios. 
 
\end{abstract}

\maketitle

\section{Introduction}
In \cite{Gross:2021gsj} the string amplitude for the decay of a highly excited string (HES) state into two tachyons was found to be an erratic function of an angle, extremely sensitive to the partition of the level $N$ associated to the choice of DDF operators generating the HES \cite{DelGiudice:1971yjh,Hindmarsh:2010if,Skliros:2011si,Skliros:2016fqs,Bianchi:2019ywd,Aldi:2019osr,Addazi:2020obs,Aldi:2020qfu,Aldi:2021zhh}. A more quantitative analysis was performed in \cite{Bianchi:2022mhs,Bianchi:2023uby}, where we showed that the extrema points  of the amplitude can be mapped into the eigenvalues of some Random Matrix Theory (RMT) for the Gaussian Unitary Ensemble (GUE) .   The map  holds for the distributions of  the spacings between successive extrema and their ratios. The ``scattering form factor" (ScFF) \cite{Bianchi:2024fsi}  exposes the same behavior as the Spectral Form Factor (SFF) of the GUE, including a ramp.  The map was demonstrated also for other chaotic  processes  unrelated to string scattering in \cite{Bianchi:2023uby} and \cite{Ageev:2025yiq}. 

In \cite{Bianchi:2025kna} we generalized the one-dimensional chaotic behavior to a higher dimensions for which  erratic functions  depend on several variables, and proposed measures for the spacings of peaks in two or more dimensions. We examined their behavior in the context of pinball scattering, and analyzed a toy model of the electric potential produced by randomly placed charges. 

Higher-dimensional chaotic behavior can be realized also for string scattering amplitudes that depend on two or more dynamical variables. This will serve as our laboratory in the present investigation.

The decay and scattering amplitudes for these cases may have a set of extrema points as for the one dimensional case but in addition may also have saddle points, curves in the form of   ``ridges", ``valleys", or other ``topographic structures". 

The goal of the present investigation is to look for curves in the scattering amplitudes and to determine their chaotic behavior by introducing a tool that  maps them to RMT. 

We define curves associated with  an erratic function  of two variables ${\cal A}(x,y)$ as  configurations  
 $y_{(x)}(x)$ and $y_{(y)}(x)$ that solve the equations
    \begin{equation}
        {\cal F}_x\big(x,y_{(x)}(x)\big) = 0\,, \qquad {\cal F}_y\big(x,y_{(y)}(x)\big) = 0 \,.
    \end{equation}
    where ${\cal F}_i(x,y)= \partial _i \log {\cal A}(x,y)$, for $i=x,y$, are the log-derivatives of the function being investigated.

We will argue that the curves associated with chaotic processes are non-intersecting. This can be phrased as a ``repulsion'' between curves. To quantify this property we need to  measure the distances between adjacent curves. A natural way to do it is by defining the areas under the curves $y_{(i)}(x)$ as:
    \begin{equation}
        A_n \equiv \int_{x_{\text{min}}}^{x_{\text{max}}} dx \,(y_n(x) - y_{\text{min}})
    \end{equation}
    Here and in the following we omit the index $(i)$ to simplify the notation, but it should be remembered that there are always two sets of curves associated with the partial derivatives with respect to the two variables. The index $n$ labels all the different curves within each set.
    
    We will be interested in the areas between adjacent curves, so we define the spacings of these ``area eigenvalues'', analyzing the distribution of the spacings and their ratios:
    \begin{equation} \label{eq:deltar}
        \delta_n \equiv A_{n+1}-A_n\,,\qquad r_n \equiv \frac{\delta_{n+1}}{\delta_n}\,.
    \end{equation}
    Note that the areas between the curves could have been computed equivalently starting from the parametrization  $x(y)$. They do not depend on the choice of a preferred variable.

    The expected distribution of the spacing ratios in RMT is \cite{Atas:2013dis}:
    \begin{equation}
    f_\beta(r) = \frac{3^{\frac{3+3\beta}2}\Gamma(1+\frac\beta2)^2}{2\pi \Gamma(1+\beta)} \frac{(r+r^2)^\beta}{(1+r+r^2)^{1+\frac32\beta}} \,,\label{eq:beta_r}
    \end{equation}
    where $\beta = 1 $ for the Gaussian Orthogonal Ensemble (GOE), 2 for the Gaussian Unitary Ensemble (GUE), and 4 for the Gaussian Symplectic Ensemble (GSE). More generally, one can define the Gaussian $\beta$-ensemble (GBE) for any value $\beta > 0$ \cite{Forrester:book}. We often use the normalized ratios $\tilde r_n \equiv \min(r_n,\frac{1}{r_n})$ which take values between 0 and 1, and whose distribution is the same as in \eqref{eq:beta_r} (with an additional factor of 2 for normalization), due to the symmetry $r \leftrightarrow \frac{1}{r}$.
    
    Another important property that will be  determined is the ``Area Form Factor" (AFF) associated the areas $A_k$ which is the analog of the ``Scattering Form Factor" of \cite{Bianchi:2024fsi}: 
    \begin{equation}
    R_2(t) \equiv \big\langle \frac{1}{L}\sum_{m=1}^{L}\sum_{n=1}^L e^{i(A_m-A_n) t} \big\rangle \,.\label{eq:sff}
\end{equation}
The brackets denote an ensemble average. After the averaging, the result can be compared to the SFF of RMT. Note that it is conventional to define the SFF with a prefactor of $1/L^2$, while here we choose a normalization with $1/L$ for later convenience.

Non-intersecting paths were discussed  in  various different systems, see for instance  \cite{KarlinMcGregor1959,Lindstrom1973,GesselViennot1989,Johansson2002}. In particular similar curves which describe the time evolution  of eigenvalues appear in the context of ``Dyson Brownian motion" \cite{Dyson1962}. The phenomenon considered in this letter is also a repulsion of curves but its origin and properties are different from those discussed in the past. In many cases, the curve repulsion is observed in the behavior of $\lambda_n(t)$, being the evolution of (one-dimensional) eigenvalues as a function of some parameter of the theory. In our example, the curves $y(x)$ will be curves in a two-dimensional plane of dynamical variables, with the two variables being treated on equal footing.

\section{The \texorpdfstring{$HES + T\to T+T$}{HES + T --> T + T} scattering amplitude}
\begin{figure}[h!]
\centering
\includegraphics[width=0.48\textwidth]{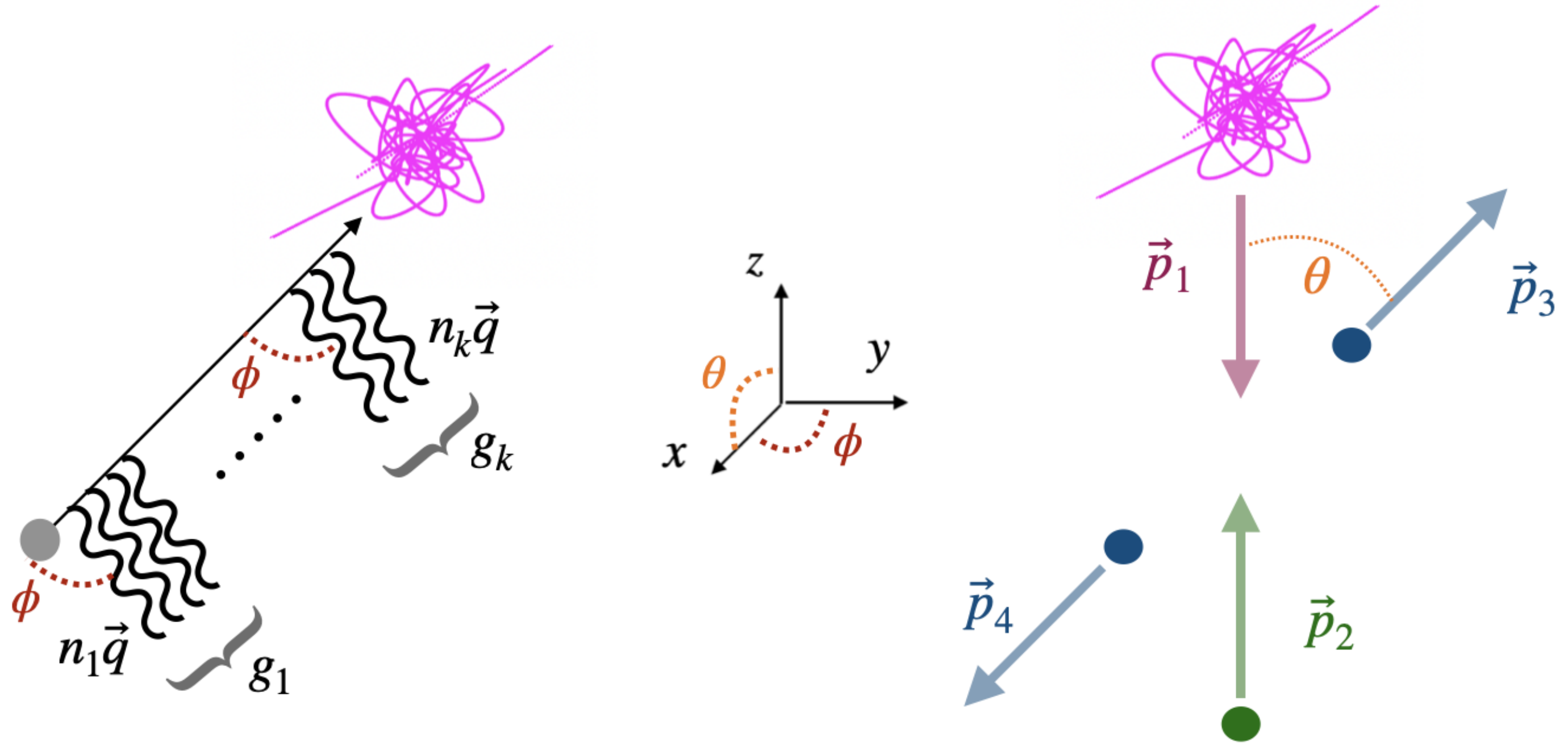}
\caption{Kinematical configuration of the formation of the DDF HES (left) and the process $HES+T\rightarrow T+T$ in the center of mass frame (right).}
\label{fig:stringConfScattkine}
\end{figure}
In \cite{Bianchi:2023uby} we analyzed a three-point function decay amplitude that depends on one angle. To discuss amplitudes that are function of two angles one has to  look at four-point functions. There are several possibilities \footnote{In a sequel paper we will analyze all processes that involve one and two highly-excited states.}.

The process we consider here is the scattering of a generic, highly-excited open bosonic string state with a massive scalar (tachyon), yielding a two-scalar final state, as represented in figure \ref{fig:stringConfScattkine}. The scattering amplitude can be efficiently computed following \cite{Firrotta:2024qel}.

A convenient Lorentz-covariant setup is provided by the following four-dimensional kinematics:
\begin{align}
&p_{1}=(\sqrt{p_{i}^{2}{+}M_{N}^{2}},p_{i}\hat{n}_{i})\,,
&p_{2}=(\sqrt{p_{i}^{2}{+}M^{2}},-p_{i}\hat{n}_{i})\,, \\ 
&p_{3}=(-\sqrt{p_{o}^{2}{+}M^{2}},p_{o}\hat{n}_{o})\,,
&p_{4}=-(\sqrt{p_{o}^{2}{+}M^{2}},p_{o}\hat{n}_{o})\,, \nonumber
\end{align}
with spatial orientations $\hat{n}_{i}=(0,0,1)$ and $\hat{n}_{o}=(\sin\theta,0,\cos\theta)$. The magnitudes of the incoming and outgoing momenta, $p_i$ and $p_o$ respectively, are expressed in terms of the mass of the highly excited state $M_N^2=(N{-}1)/\alp$, the mass of scalar $M^2=-1/\alp$ and the Mandelstam variable $s\equiv-(p_1{+}p_2)^2=-(p_3{+}p_4)^2$ through
\begin{equation}
p_{i}={\sqrt{[s{-}(M_{N}{+}M)^{2}][s{-}(M_{N}{-}M)^{2}]}\over 2\sqrt{s}}\,, p_{o}=\sqrt{{s\over 4}{-}M^{2}}.
\end{equation}
The complexity of the HES is parametrized in terms of the reference null momentum $q^\mu \equiv q_r^{\mu}/(2\alp q{\cdot}p_1)$ parallel to $p_1$, with
\begin{equation}
q^{\mu}_{r}={(1,\cos\phi,\sin\phi,0,\vec{0})}
\end{equation}
and the reference polarization $\zeta^\mu \equiv \lambda^{\mu}{-}\lambda{\cdot}p q^\mu$ with
\begin{equation}\label{polvec}
\lambda^{\mu}=(0,-\sin\phi,\cos\phi,0,\vec{0})
\end{equation}
providing the physical polarization tensor
\begin{equation}
H_{(\mu_1)_{g_1}...(\mu_N)_{g_N}} \equiv \bigotimes_{n=1}^N\bigotimes_{i=1}^{g_n}\zeta^{(n)}_{\mu_i}
\end{equation}
transverse to the HES momentum $p^\mu_1$ and diagonal in the basis of string partitions at excitation level $N$, $N=\sum_{n=1}^N n g_n$,
where $n$ identifies the specific harmonic present in the string state, in general excited $g_n$  times.

To gain deeper insight into the complexity of the HES, we analyze the short-distance regime defined by fixed-angle high-energy scattering, in which $s$ and $t \equiv -(p_1+p_4)^2$ are taken to be large while keeping the ratio $t/s$ fixed \cite{Gross:1987ar}. The resulting scattering process is fixed by positivity, unitarity and scattering duality \cite{Firrotta:2024qel,Firrotta:2024fvi} and up to irrelevant constants, the leading contribution takes the following form \footnote{Although subleading corrections of order ${\mathcal O}(1/\alp s)$ and ${\mathcal O}(1/\alp t)$ are present, they are further suppressed by the mass of the HES and therefore do not modify the qualitative behavior of the process.}:
\begin{equation}\label{scatamp}
{\cal A}^{\{n,g_n\}}_{f.a.}(s,\theta,\phi)=g_{\text{open}}^2\,e^{-\alpha^{\prime} s f(\theta)} {\cal D}_{\{n,g_n\}}(\theta,\phi)\,,
\end{equation}
where $g_{\text{open}}$ is the coupling of open strings.
The exponential falloff, modulated by the function
\begin{equation}
f(\theta)=\bigg|\sin^2{\theta\over 2}\log\left( \sin^2{\theta\over 2}\right)+\cos^2{\theta\over 2}\log\left( \cos^2{\theta\over 2}\right)\bigg|
\end{equation}
is the characteristic feature of high-energy string processes and reflects the softness of string interactions at short distances. The factor $e^{-\alpha^\prime s f(\theta)}$ is precisely the form of the Veneziano amplitude in the high-energy fixed-angle limit.

In the case of HES, however, this behavior is further accompanied by the nontrivial stringy dressing factor
\begin{equation}\label{DressPart}
    {\cal D}_{\{n,g_n\}}(\theta,\phi)= \left({\sqrt{\alpha^{\prime}s}\sin\phi \sin\theta\over \cos^{2}{\theta\over 2}} \right)^{J} \prod_{n=1}^{N}\big({\cal P}_n(\theta,\phi) \big)^{g_{n}}\,,
\end{equation}
which encodes information about the underlying complexity of the process. This complexity originates from the structure of the HES and is manifested through the highly nonlinear, partition-dependent combination of Jacobi polynomials:
\begin{equation} \label{eq:polyn}
    {\cal P}_n(\theta,\phi) =P_{n-1}^{(-2n+1,-n\sin\theta\cos\phi)}\left(1{-}\frac{2}{\cos^{2}\frac\theta2}\right)\,,
\end{equation}
whose intricate product representation generates a rich two-dimensional angular profile for the scattering amplitude. This profile depends non-trivially on both the scattering angle $\theta$ and the orientation angle $\phi$ characterizing the HES.

The highly nonlinear and partition-dependent structure of \eqref{DressPart} points toward a previously unexplored link between the microscopic complexity of HES and the emergence of a higher-dimensional chaotic behavior. This remarkable structure provides a unique opportunity to investigate whether such complexity leaves measurable imprints on scattering observables and, more broadly, to understand how chaos arises in string dynamics. Motivated by these questions, we start a quantitative analysis of the HES contribution and probe it for distinctive signatures of chaos in higher dimensions.

\section{Analysis of the amplitude and its curves}
To carry out the analysis, we rewrite  \eqref{scatamp} in terms of $x=\cos\theta$ and $y=\cos\phi$, determine numerically all the curves $y_{(i),n}(x)$ for each partition in a given sample of HES states, and use them to diagnose the chaotic behavior.

We focus only on the angle-dependent part of the dressing factor \eqref{DressPart}. To avoid cumbersome terminology, we refer to it henceforth simply as ``the amplitude''. In \cite{Bianchi:2023uby}, we considered also the effect of the Veneziano factor in \eqref{scatamp}, and showed that, when the polarization angle is fixed, the full amplitude as a function of the scattering angle displays a transition from regular spacings of the peaks at large angles to chaotic GUE spacings at smaller angles, depending on which term dominates.

\begin{figure}[ht!]
        \centering
        \includegraphics[width=0.48\textwidth]{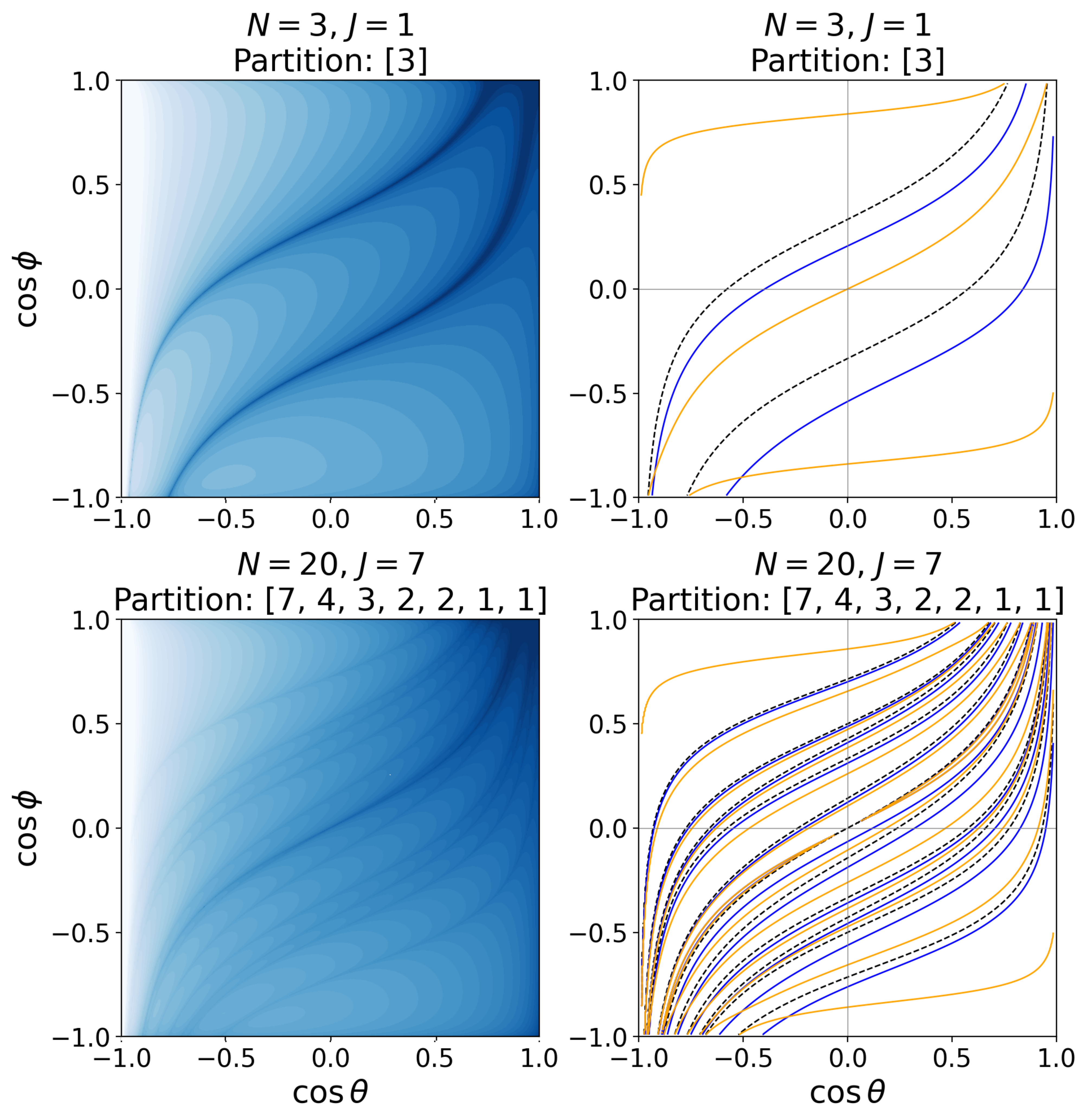}
         
        \caption{Amplitudes of two states at $N = 3$ and 20 and their curves. Left: contour plot of $\log|{\cal D}(x,y)|$). Darker regions represent smaller values. Right: plot of the curves of zeros of the amplitude (dashed, black) and its $x$- and $y-$derivatives (blue and orange, respectively).}
        \label{fig:amp23}
    \end{figure}

In figure \ref{fig:amp23}, we plot the dressing factor as a function of $(x,y) = (\cos\theta,\cos\phi)$ for two choices of states at levels $N = 3$ and $N = 20$. At all levels we see a similar structure of the amplitude. There are curves along which the amplitude vanishes, which are determined by the zeros of the Jacobi polynomials that appear in \eqref{eq:polyn}. Between each pair of zeros, there are curves corresponding to the locations of zeros of the log-derivatives of ${\cal D}(x,y)$. There is also a set of local maxima of $|{\cal D}(x,y)|$ where both derivatives vanish.

As we increase $N$, we observe that the same overall structure remains, with a higher number of curves that grows linearly with $N$. Crucially, for generic partitions of large $N$, the spacings between curves become erratic.

\begin{figure}[t!]
    \centering
    \includegraphics[width=0.48\textwidth]{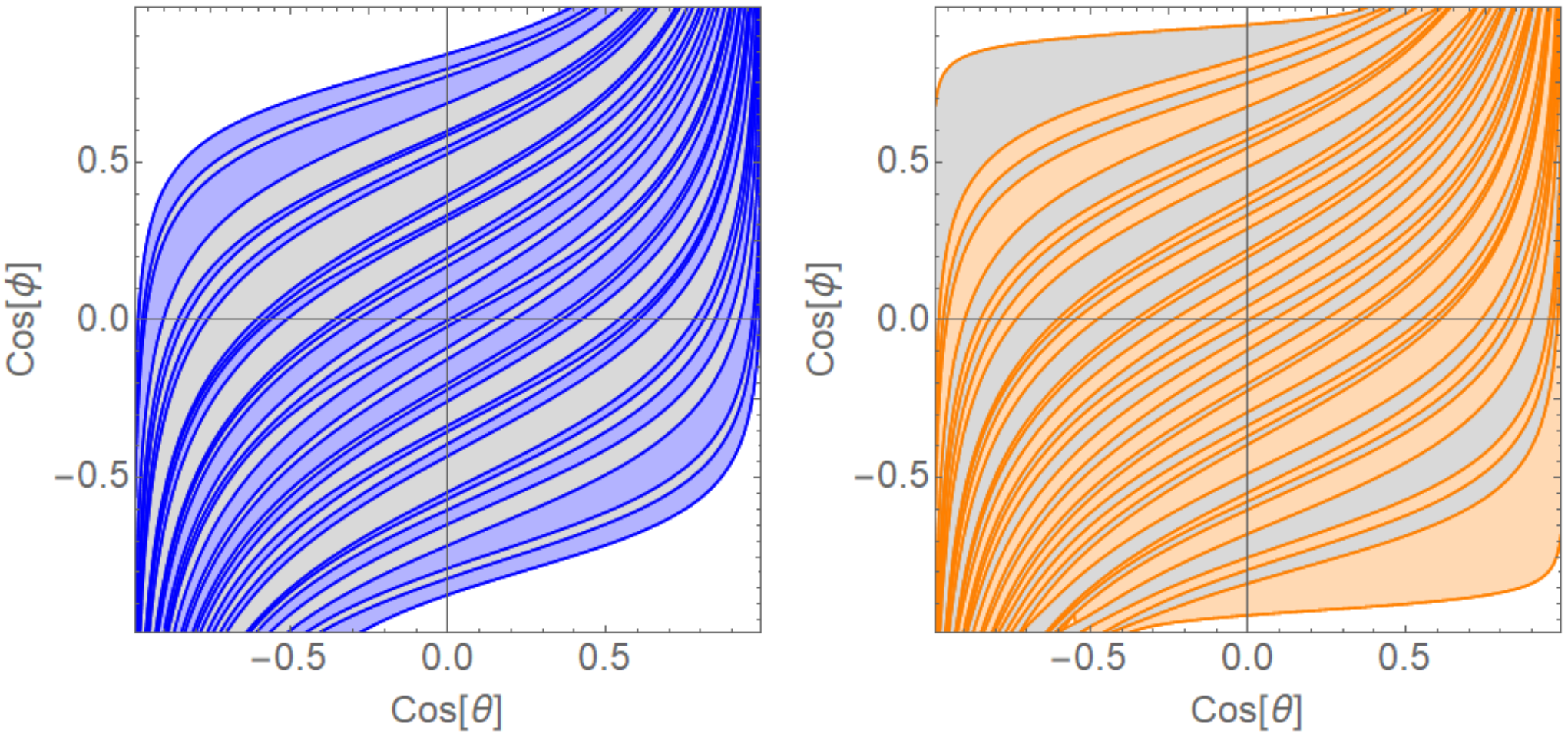}

    \caption{Curves of zeros of the $x$-derivative (left), and $y$-derivative (right) of $\log{\cal D}(x,y)$, for the state with $N = 50 = 13+10+9+5+5+3+2+1+1+1$. The areas colored in alternating colors are the spacings $\delta_n$.
    }
    \label{fig:areas}
\end{figure}

In figure \ref{fig:areas}, we plot the curves for the a randomly selected HES with $N=50$. The colored areas between the curves in the figure correspond to the ``level spacings'' $\delta_n$ whose distributions we will analyze below. One can observe immediately that the curves are non-intersecting and one can view this phenomenon as curve repulsion.

\subsection{Spacing ratios}
To carry out a quantitative analysis, we determine the two types of curves over samples of 100 states at each level $N$, chosen by taking random integer partitions of $N$. Then, we take all the resulting spacing ratios for the areas, $\{r_{(x),n}\}$ and $\{r_{(y),n}\}$, and fit their distributions to the $\beta$-ensemble formula \eqref{eq:beta_r}. Independently of the fitting, we can compute the average value of the normalized spacing ratios $\langle \tilde r_n\rangle$, and compare it to the expected values for the non-chaotic Poisson distribution (0.386) and the expected RMT values of 0.536 for GOE and 0.603 for GUE. The results are shown in table \ref{tab:fitvalues}.

\begin{table}[h!]
    \centering
    \begin{tabular}{c|c|c|c|c|c}\
     N\,  &  40 & 50 & 60 & 80 & 100 \\ \hline
     $\beta_{(x)}$\, & 1.36  & 1.45 & 1.24 & 1.22 &  1.33 \\
     $\beta_{(y)} $\, & 3.45 & 3.01 & 2.81 & 2.69 & 2.63 \\ \hline
     
    $\langle {\tilde r}_{(x)}\rangle$\, & 0.547  & 0.551 & 0.538 & 0.533 &  0.546 \\
    $\langle {\tilde r}_{(y)}\rangle$\, & 0.685 & 0.662 & 0.653 & 0.640 & 0.643 \\ \hline
     
    \end{tabular}
    \caption{Values of fitted $\beta$ and average of $\tilde r_n$ for the two types of curves, as a function of the level $N$.}
    \label{tab:fitvalues}
\end{table}

The distributions obtained are plotted in figure \ref{fig:areas_spacings} for the case of $N = 100$. The rest of the distributions can be seen in Appendix \ref{app:figures_rn} of the Supplemental Material. It should be noted that at larger $N$, the distributions agree better with the RMT formula. For $N = 100$, the best fit for the distribution computed from the curves where the $x$-derivative vanishes has $\beta=1.33$, while for the $y$-derivatives we find $\beta=2.63$.

This is an intriguing result. The angle $\phi$ (corresponding to $y$) is related to the polarization vector \eqref{polvec}. This is similar to the angle that appeared in the decay associated with a three-point function \cite{Bianchi:2022mhs}. In that case the results admitted an extrapolation to $\beta\approx2$, as in the GUE. On the other hand $\theta$ (corresponding to $x$), which is a genuine scattering angle, seems to be invariant under a $Z_2$ symmetry and hence, as in quantum chaotic systems with time-reversal invariance, is related to the GOE distribution with $\beta=1$. We can also observe that whereas  $\beta_{(y)}$ and $\langle\tilde r_{(y)}\rangle$ are monotonically decreasing their analogs in $x$ oscillate. It is not clear why that is the case.

\begin{figure}[h!]
    \centering

    \includegraphics[width=0.23\textwidth]{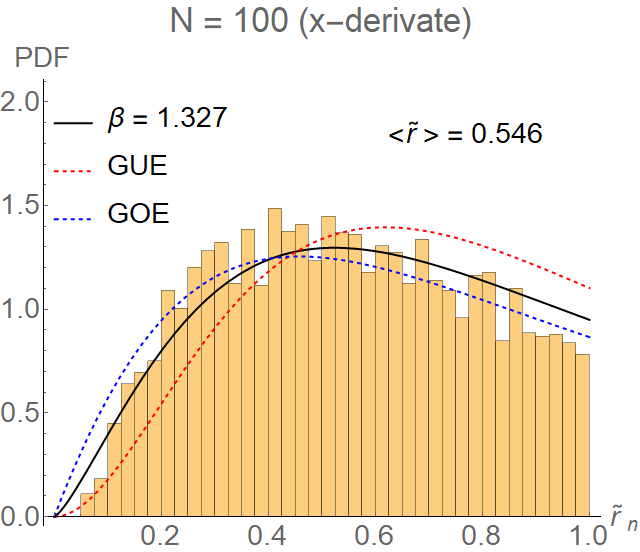}
    \includegraphics[width=0.23\textwidth]{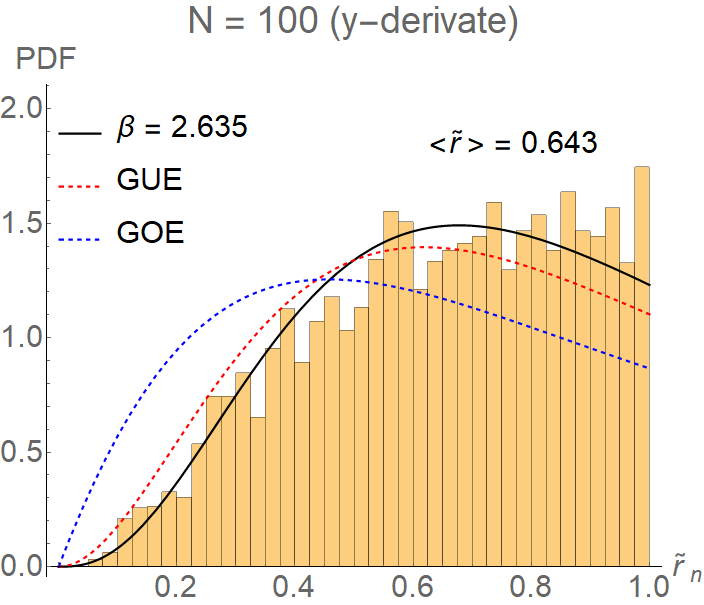}
    \caption{Distributions of ratios of adjacent areas compared to RMT (eq. \eqref{eq:beta_r}) for the GOE, GUE, and best-fitted $\beta$ values. The distribution is averaged over 100 states of $N = 100$. The left and right plots correspond to the areas associated with the $x$- and $y$-derivatives, respectively.
    }
    \label{fig:areas_spacings}
\end{figure}
\vspace{-0.8cm}

\subsection{Area Form Factor}
To provide an additional verification of the correspondence of the distribution of the area eigenvalues to RMT, we compute the ``area form factor", defined in eq. \eqref{eq:sff}. Note that in our ``ensemble average'', the size $L$ is a variable, as the number of eigenvalues depends on the details of the partition. To resolve this, we use the following normalization. First, the areas are by definition $A_n\in[0,4]$. We rescale all areas $A_n \to \tilde A_n$ such that the average difference in the rescaled variable, $\langle \tilde A_{n+1}-\tilde A_n\rangle$, equals one for each of the sets of eigenvalues. The AFF can then be plotted as a function of the time variable $\tau = t/2\pi$, in terms of which the plateau region is expected to begin at $\tau = 1$.

Then, we perform the ensemble averaging after normalizing the AFF such that the value of the AFF at the late-time plateau is one. This is the reason for the factor of $\frac{1}{L}$ in eq.~\eqref{eq:sff}.

In addition, the AFF, as the SFF, can be separated into the disconnected and connected pieces of the two-point correlator. The disconnected piece is essentially the Fourier transform of the average eigenvalue density function. To compare our results directly to the universal RMT predictions, we subtract the disconnected piece numerically by computing:
\begin{equation} \label{eq:sffconn}
    R_2^{\mathrm{conn.}}(t) =  R_2(t) - \big\vert\big\langle \frac{1}{\sqrt{L}} \sum_{n=1}^L e^{i\lambda_n t} \big\rangle\big\vert^2\,,
\end{equation}
where the eigenvalues $\lambda_n$ are taken to be the rescaled areas $\tilde A_n$. The factor of $1/\sqrt{L}$ inside the average is chosen for consistency with our normalization. 

The connected piece $R_2^{\mathrm{conn.}}(\tau)$ can be compared to the RMT prediction, which for large matrices is independent of $L$. 

\begin{figure}[h!]
    \centering

    \includegraphics[width=0.23\textwidth]{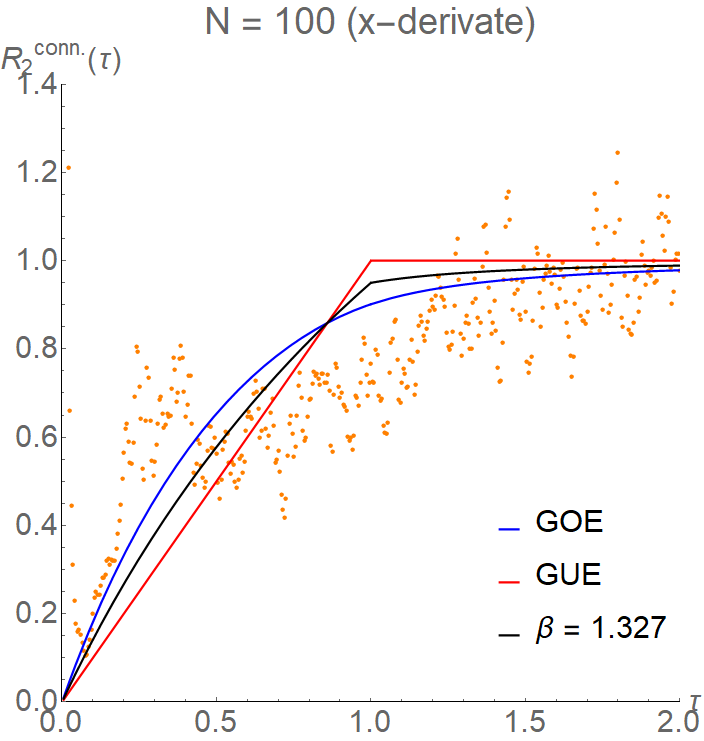}
    \includegraphics[width=0.23\textwidth]{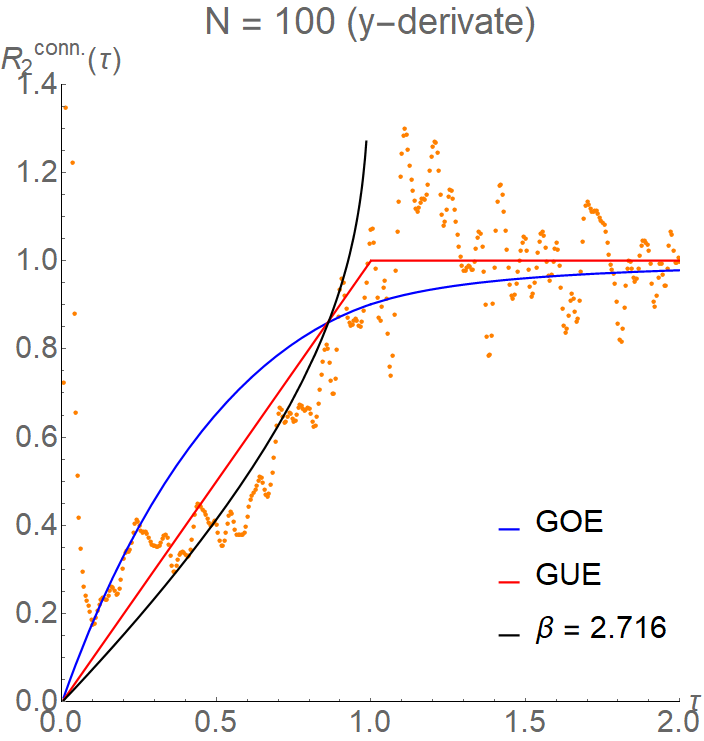}
    \caption{Connected part of the AFF, computed using the areas under curves as eigenvalues, for 100 randomly chosen partitions of $N = 100$.
    }
    \label{fig:areas_sff_100}
\end{figure}

For general $\beta$, we can use an interpolation between the GOE ($\beta=1$) and GUE ($\beta=2$) that we have verified numerically in \cite{Bianchi:2024fsi} (see also \cite{Weber:2020ytu}):
\begin{equation} R_2^{\mathrm{OU}}(\tau) = \begin{cases} \frac{2}{\beta}\tau - \left(\frac{2}{\beta}-1\right)\tau \log(1+2\tau) & \tau < 1 \\
\frac{2}{\beta} - \left(\frac{2}{\beta}-1\right)\tau \log\left(\frac{2\tau+1}{2\tau-1}\right) & \tau \geq 1 \end{cases} \label{eq:sff_beta}
\end{equation}
For $2\leq \beta \leq 4$, the ramp region $\tau < 1$ can likewise be described by an interpolation between the GUE and GSE ($\beta = 4$) formulae \footnote{Note that the interpolating formula for $2 < \beta\leq4$ has a logarithmic singularity at $\tau = 1$. This is a feature of the GSE, but not of $\beta < 4$. Because of this, the interpolation is not reliable at $\tau = 1$, and it is plotted for $\tau \leq 0.985$ in figure \ref{fig:areas_sff_100}.},
\begin{equation}
 R_2^{\mathrm{US}}(\tau) = \frac{2}{\beta}\tau -\frac12\left(1-\frac{2}{\beta}\right) \tau \log|1-\tau| \label{eq:sff_beta2}
\end{equation}

The results for $N = 100$ are given in figure~\ref{fig:areas_sff_100} \footnote{Additional figures for other values of $N$ are in Appendix~\ref{app:figures_sff} of the Supplemental Material.}. They show a clear ramp to plateau structure, as expected from RMT. The results of the AFF are largely consistent with the values of $\beta$ which we extract from the fits of the spacing ratios in table~\ref{tab:fitvalues} (note that $\beta$ is not fitted again independently here). In particular, they provide additional evidence of a qualitatively different behavior of the $x$ and $y$ sets of eigenvalues. The former display a concave ramp consistent with $1\leq \beta < 2$ while the latter is convex, as is the case for $\beta > 2$.

\section{Summary  and outlook}

The landscape of multi-dimensional chaotic processes admits a rich spectrum of structures in particular extrema points, saddle points, ridges, valleys, and more. In this note we analyzed the properties of curves in  scattering amplitudes which are functions of two angles. We showed that they are chaotic in the sense that they repel each other and hence non-intersecting. More importantly, their statistics can be mapped into the $\beta$-ensemble of RMT.

The map is based on treating the areas under each curve $A_n$ as the ``eigenvalues''. We showed that the ratios between the spacings of adjacent areas and the corresponding form factor behave similar to those of the $\beta$-ensembles. We have found that the string scattering amplitude encodes two different ensembles in a single function: with GOE-like behavior for one set of curves, and GUE-like for the other.

There are several directions for future research, including:
\begin{itemize}
\itemsep0em 
\item In \cite{Bianchi:2025kna}, in the context of multi-dimensional chaotic processes, we have  identified and investigated the structure of the extrema points, rather than curves. One can ask if there are cases of string scattering amplitudes that admit extrema points distributed randomly on the plain, and what are the conditions for them to show up.
\item 
Repelling curves likely appear also in other chaotic processes, which should be identified. In string theory, the most obvious candidates are the four-point amplitudes of $HES\rightarrow T+T+T$ and $HES+T\rightarrow HES+T$.
\item
In higher-dimensional chaotic behavior there could be repelling surfaces. Detecting them and determining  whether their behavior is similar to that of the curves is also a future goal. 
\item 
As was mentioned in the introduction, repelling curves show up in other chaotic cases, for instance in \cite{Johansson2002} and \cite{Dyson1962}. It will be interesting to use the areas for their analysis. 
\item 
In \cite{Bianchi:2025kna} we conjectured about a possible relation between the multi-dimensional chaos and Random Tensor Theory (RTT). Here, by using the areas we transferred the data of the curves into point eigenvalues that were mapped to RMT. This procedure may be related to projecting eigenvalues of RTT to those of RMT. 
\item Another interesting aspect of the potential chaotic behavior of HES is related to the one-loop corrections to their masses. An initial step in this direction has been taken in \cite{Grimaldi:2026zgv,Bianchi:2026wts}.

\end{itemize}
\section*{Acknowledgments}
We thank D. Ageev, P. Di Vecchia, A. Gaikwad, K. Papadodimas, G. Parisi,  N. Shrayer, E. Kiritsis and G. Veneziano for useful comments and discussions.

MB would like to thank the MIUR PRIN contract 2020KR4KN2 ``String Theory as a bridge between Gauge Theories and Quantum Gravity"  for partial support. MB and DW thank the the INFN project ST\&FI ``String Theory and Fundamental Interactions'' for partial support. MF would like to thank CERN for its hospitality during the completion of part of this work. The work of JS was supported in part by a grant 01034816 “String theory reloaded- from fundamental questions to applications” of the “Planning and budgeting committee”. DW was supported by an INFN postdoctoral fellowship.

\textbf{Data and code availability:} The code associated with this letter is available online at \cite{stringchaosgithub:2606}.

\bibliography{SACS}

\newpage
\onecolumngrid
\appendix
\centerline{\bf \Large Supplemental Material}
\setcounter{equation}{0}
\renewcommand{\theequation}{I.\arabic{equation}}

\section{Additional figures: Spacing ratios} \label{app:figures_rn}
\begin{figure}[ht!]
    \centering
    
\includegraphics[width=0.30\textwidth]{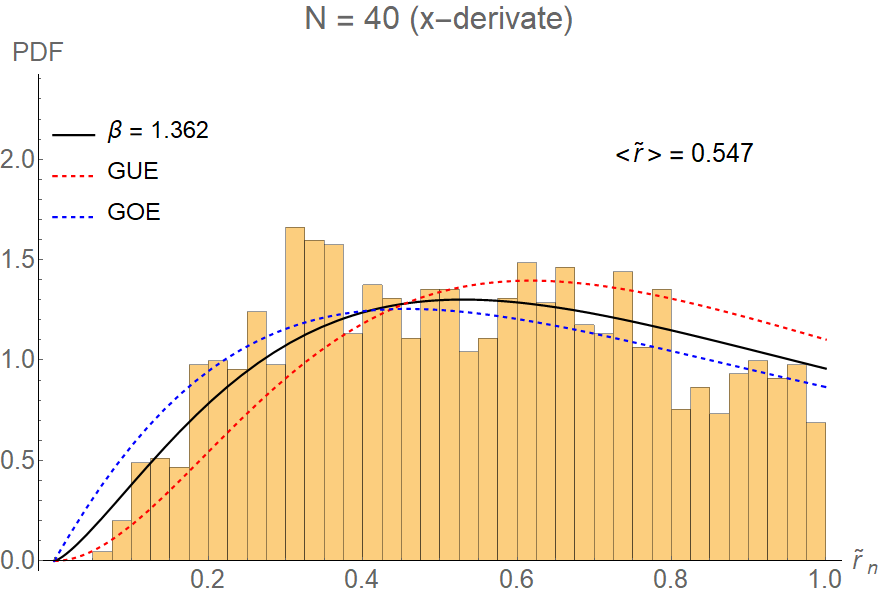}
\includegraphics[width=0.30\textwidth]{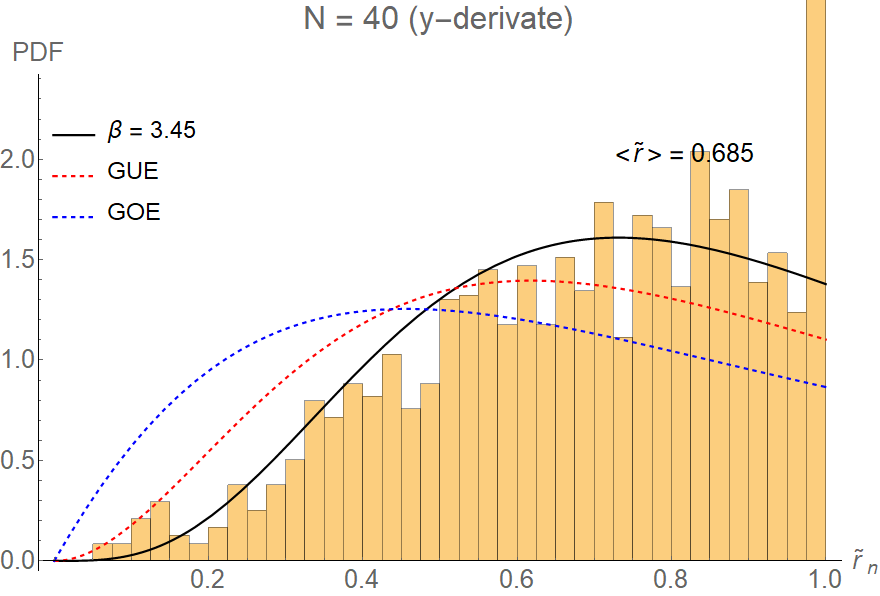} \\

\includegraphics[width=0.30\textwidth]{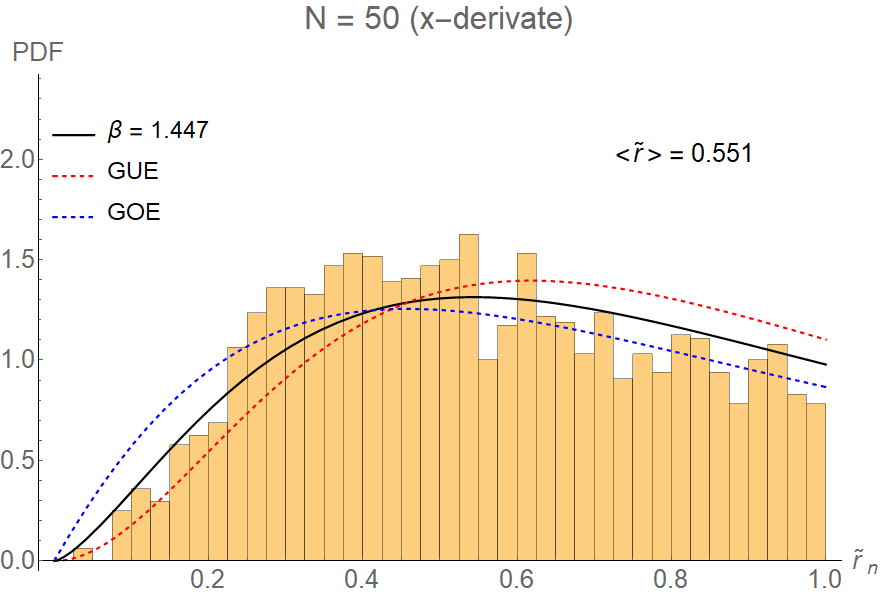}
\includegraphics[width=0.30\textwidth]{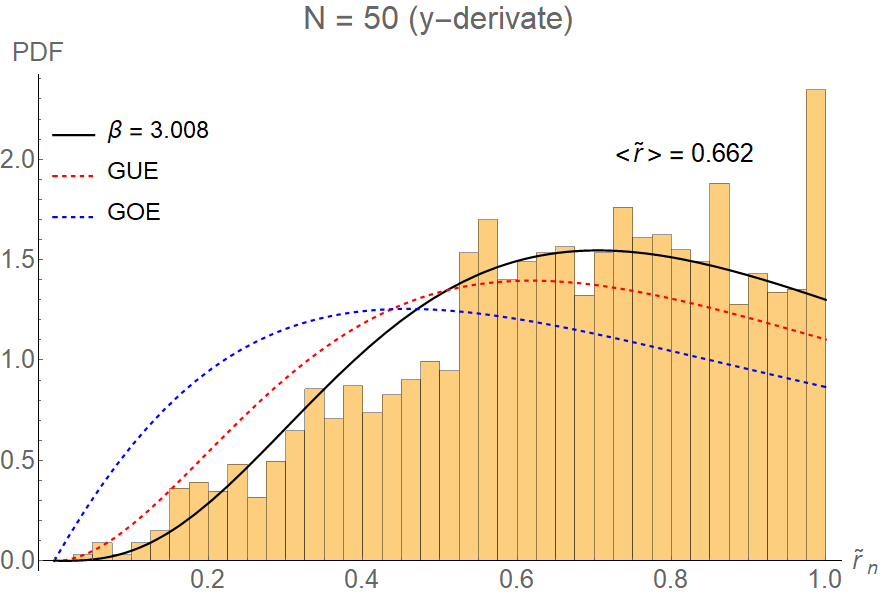} \\

\includegraphics[width=0.30\textwidth]{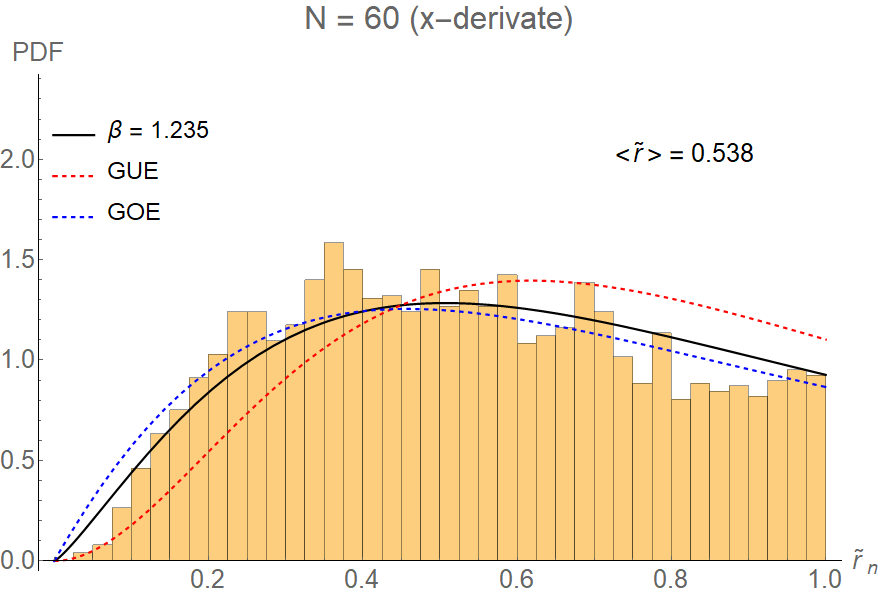}
\includegraphics[width=0.30\textwidth]{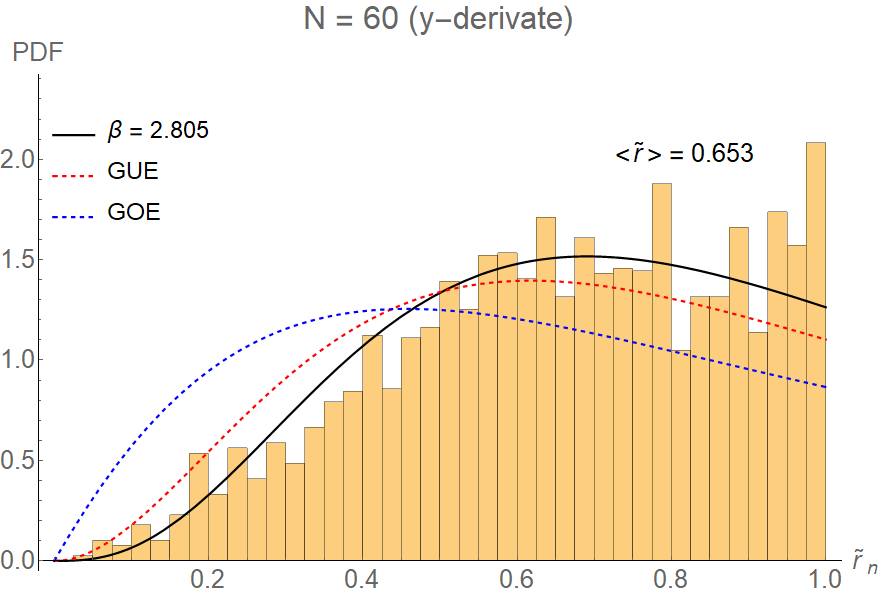} \\

\includegraphics[width=0.30\textwidth]{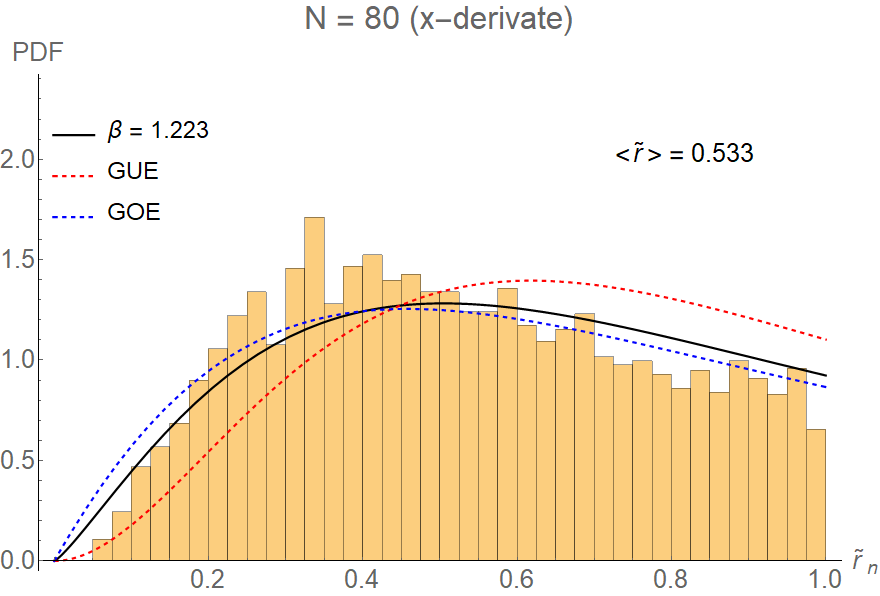}
\includegraphics[width=0.30\textwidth]{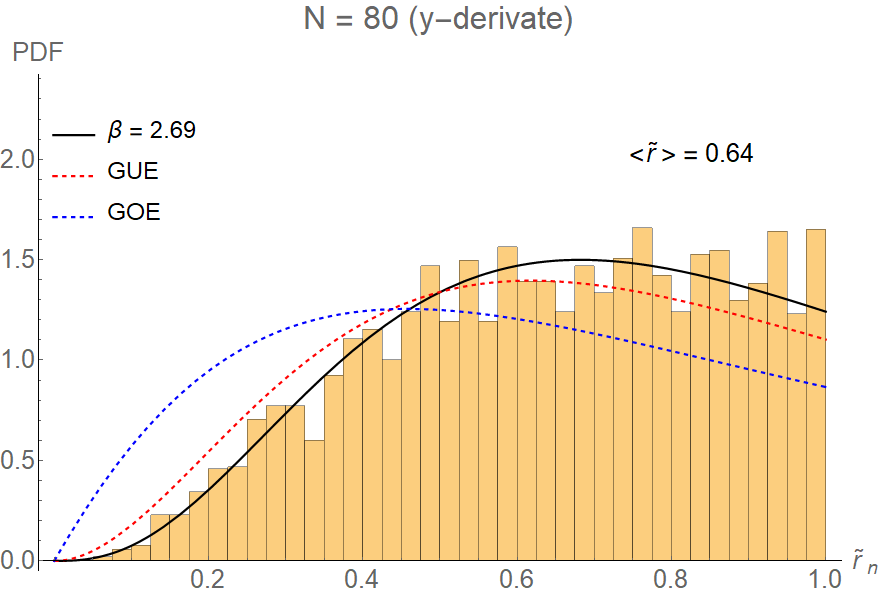} \\

\includegraphics[width=0.30\textwidth]{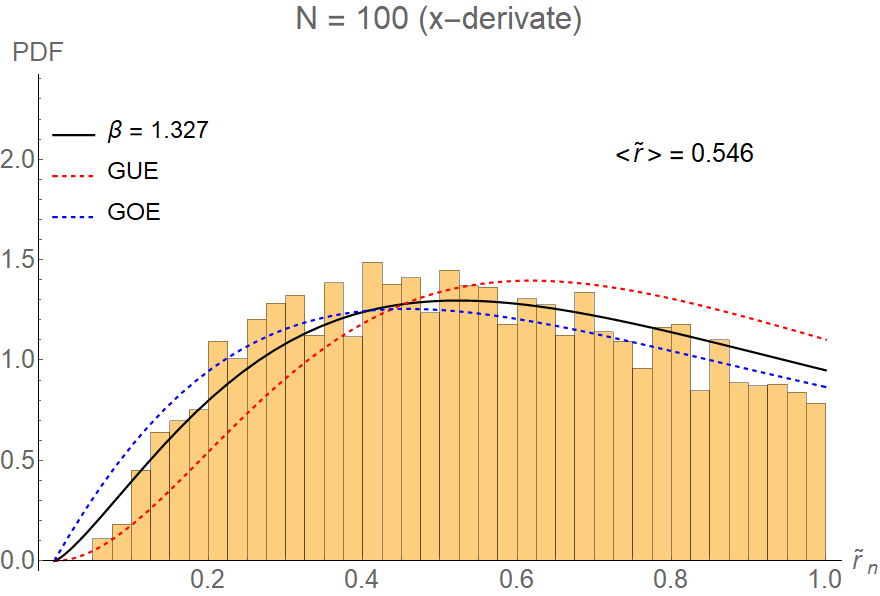}
\includegraphics[width=0.30\textwidth]{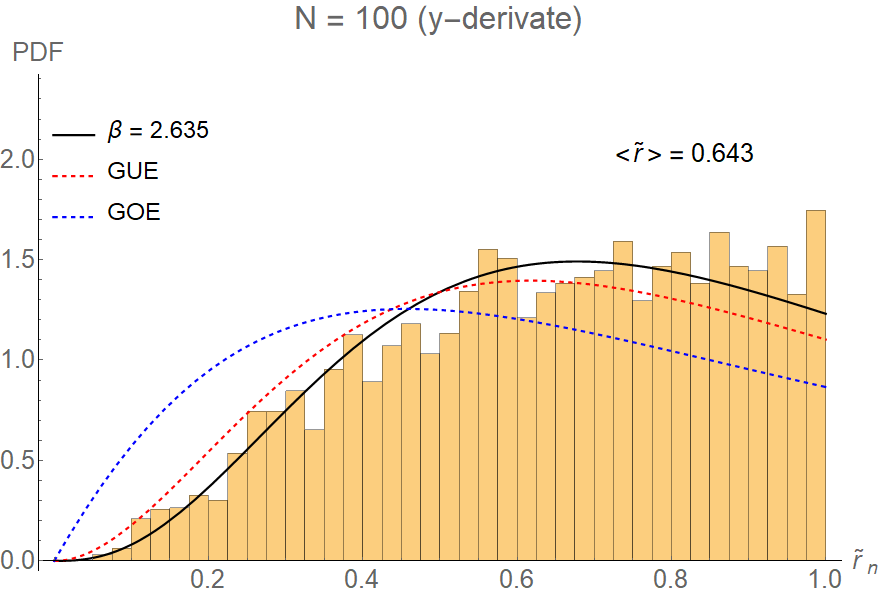} \\

    \caption{Distributions of ${\tilde r}_n$ for $N = 40$, 50, 60, 80, and 100, with best fitted value of $\beta$. The left and right columns show the distributions associated with the $x$- and $y$-derivatives, respectively.}
    \label{fig:alldistributions}
\end{figure}

In figure \ref{fig:alldistributions} are plotted the distributions of the normalized spacing ratios ${\tilde r}_n$ of the areas, computed numerically for samples of 100 random partitions at each level $N$. The results were presented in table \ref{tab:fitvalues}.

\clearpage
\section{Additional figures: Area form factor} \label{app:figures_sff}
\begin{figure}[h!]
    \centering
    
\includegraphics[width=0.30\textwidth]{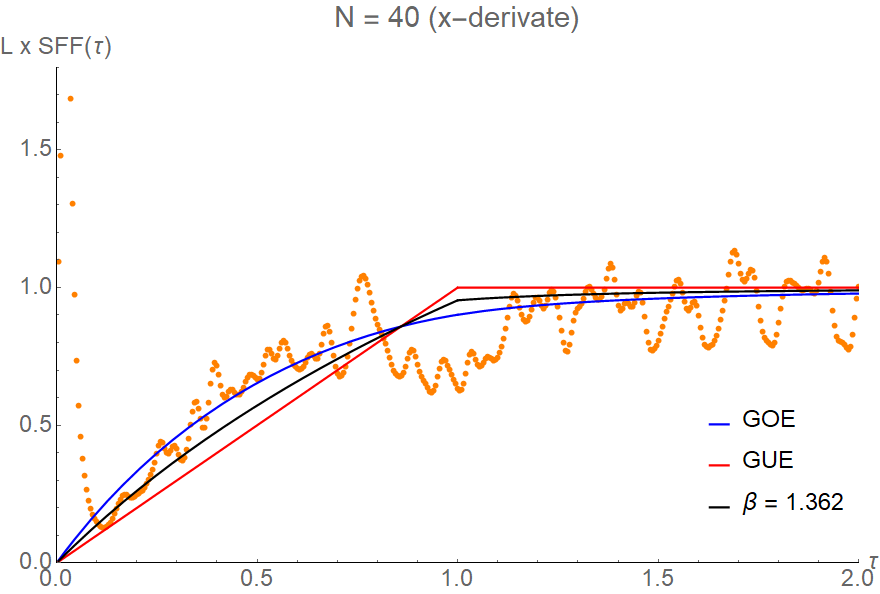}
\includegraphics[width=0.30\textwidth]{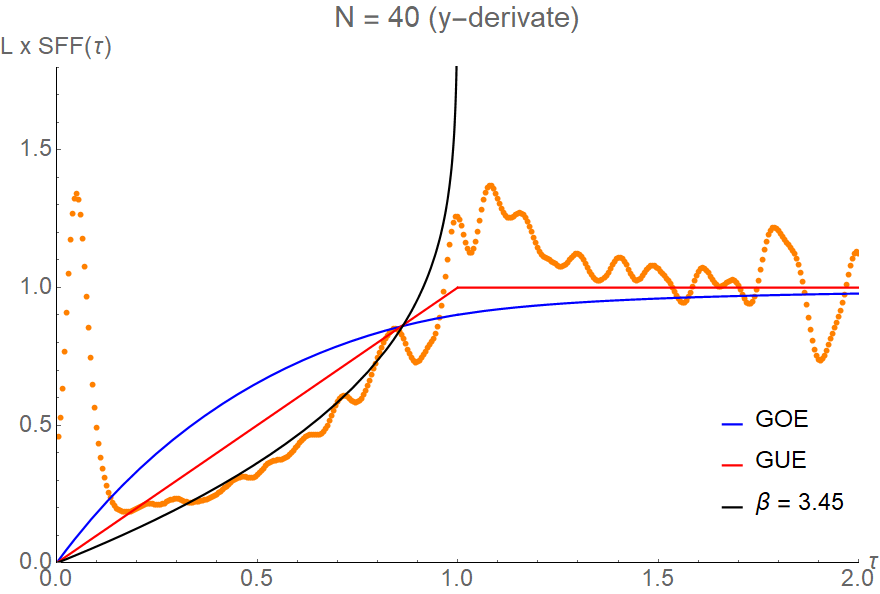} \\

\includegraphics[width=0.30\textwidth]{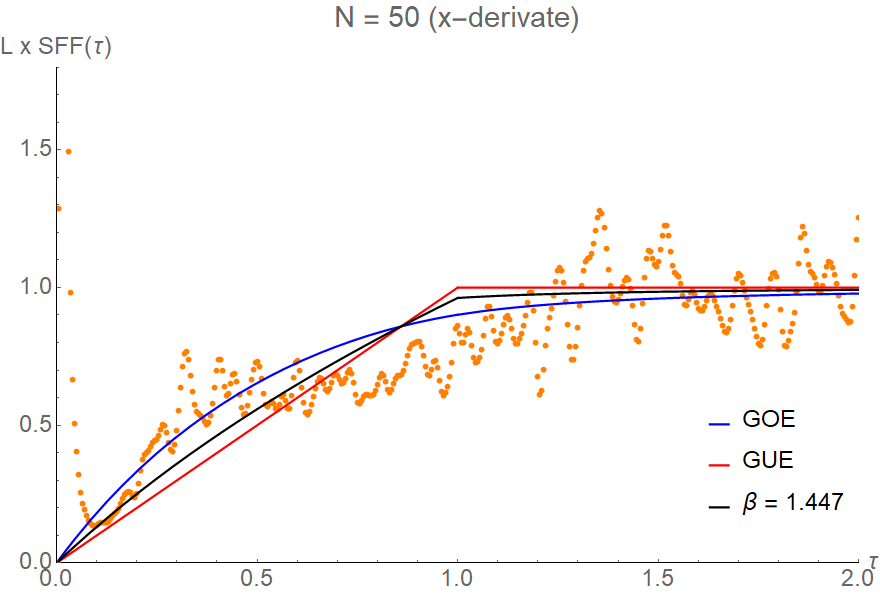}
\includegraphics[width=0.30\textwidth]{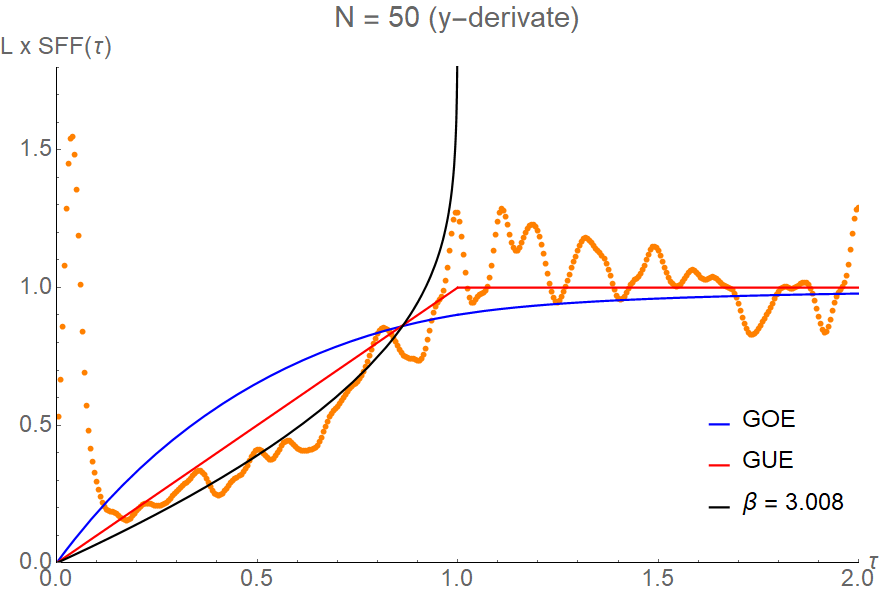} 

\includegraphics[width=0.30\textwidth]{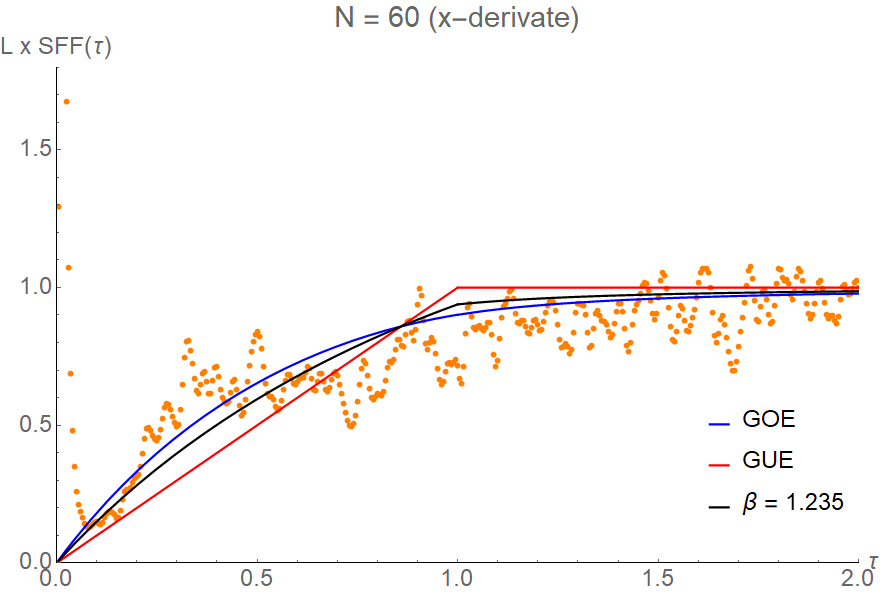}
\includegraphics[width=0.30\textwidth]{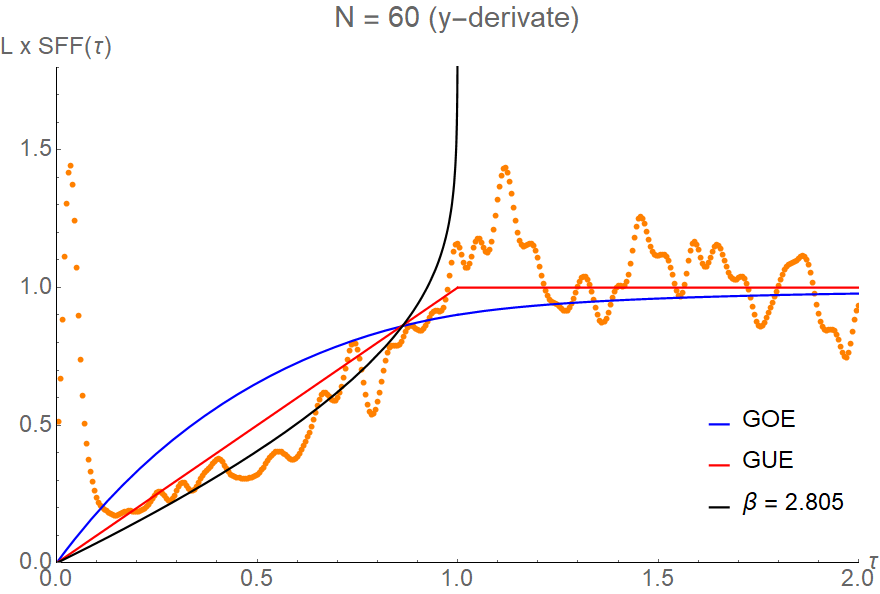} 

\includegraphics[width=0.30\textwidth]{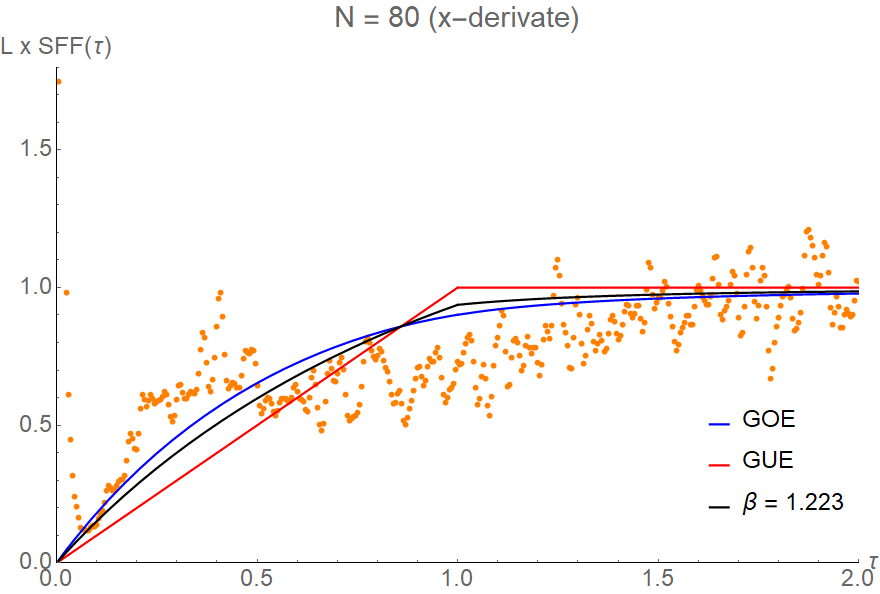}
\includegraphics[width=0.30\textwidth]{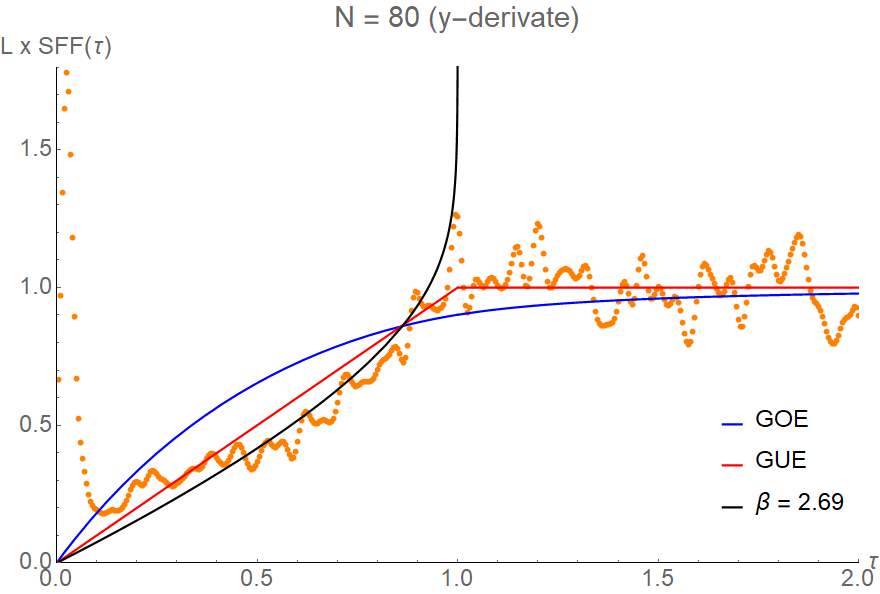} 

\includegraphics[width=0.30\textwidth]{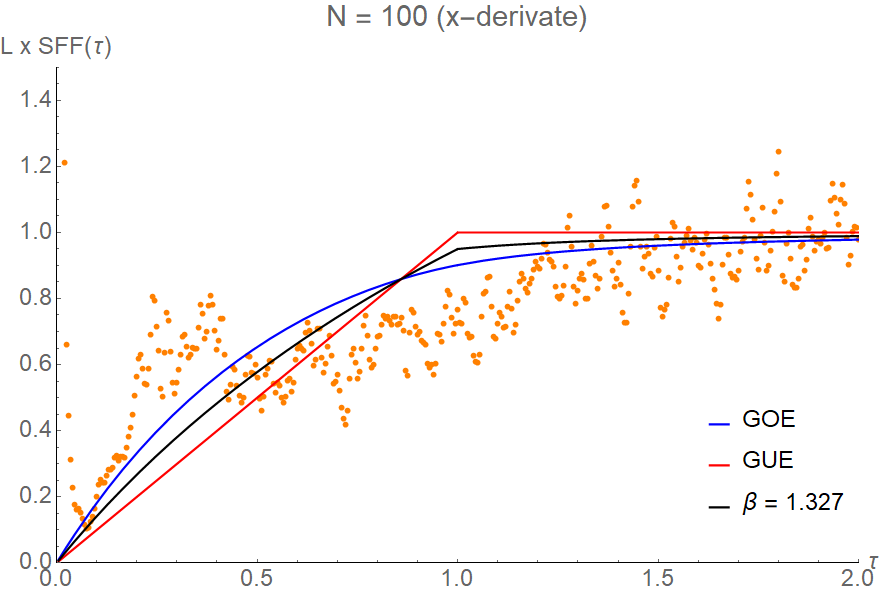}
\includegraphics[width=0.30\textwidth]{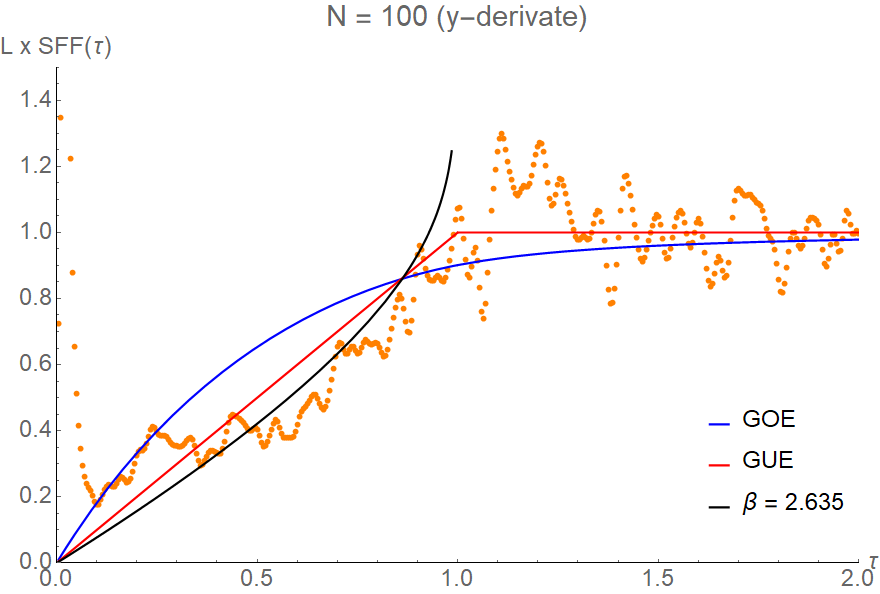} 

    \caption{Connected part of the AFF on a linear scale for $N = 40$, 50, 60, 80, and 100. The left and right columns show the distributions associated with the $x$- and $y$-derivatives, respectively.}
    \label{fig:allsfflin}
\end{figure}

In this appendix, we plot all the results for the AFF. In figure \ref{fig:allsfflin} we plot all the results for the connected part $R_2^{\text{conn.}}(\tau)$ (eq.~\eqref{eq:sffconn}) of the AFF for the different levels $N$, with the comparison to the $\beta$-ensemble formulae of eqs.~\eqref{eq:sff_beta}--\eqref{eq:sff_beta2}. The values of $\beta$ are extracted from the distributions plotted in figure \ref{fig:alldistributions}.

\begin{figure}[ht!]
    \centering
    
\includegraphics[width=0.36\textwidth]{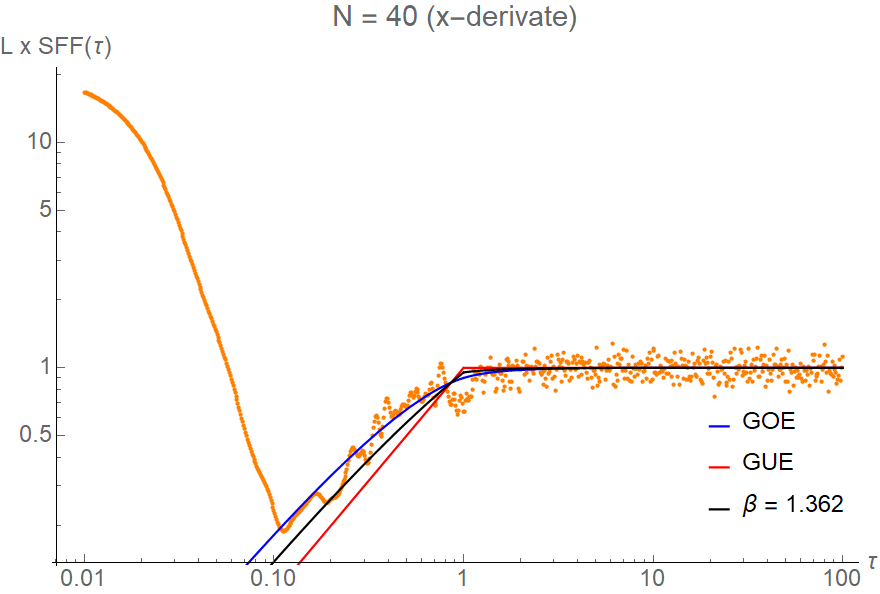}
\includegraphics[width=0.36\textwidth]{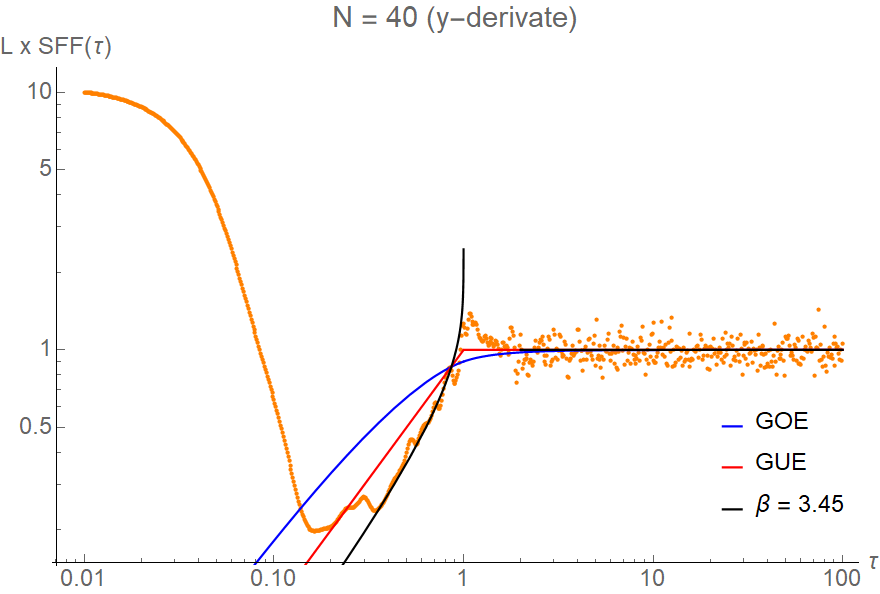} \\

\includegraphics[width=0.36\textwidth]{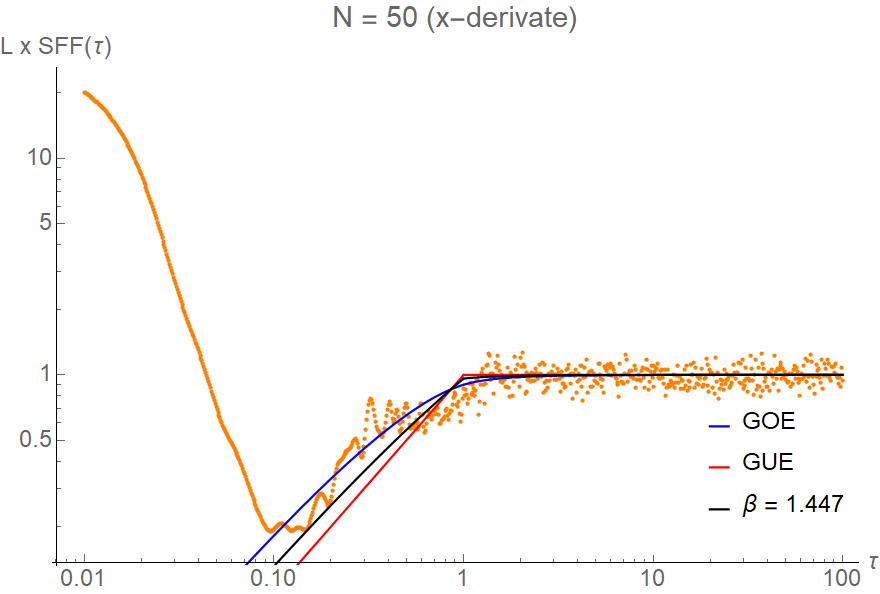}
\includegraphics[width=0.36\textwidth]{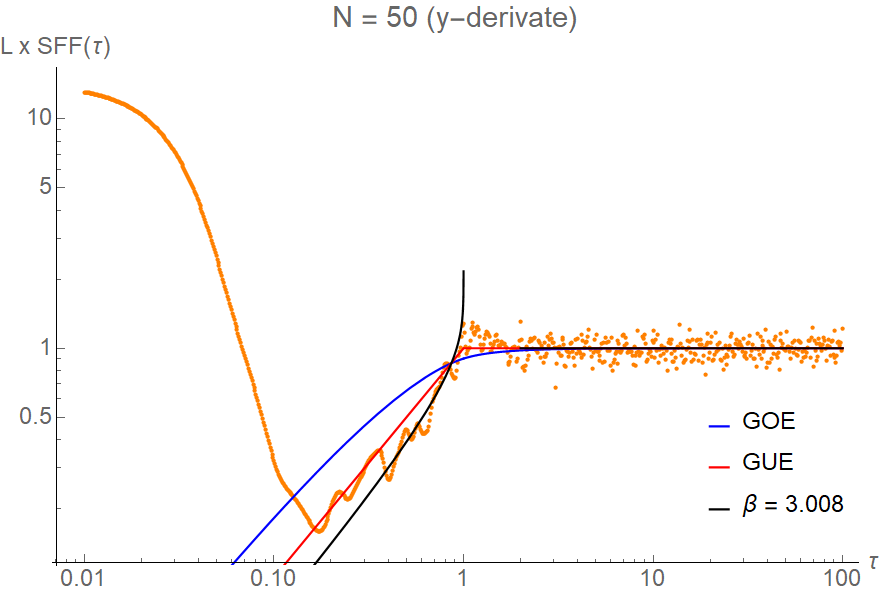} 

\includegraphics[width=0.36\textwidth]{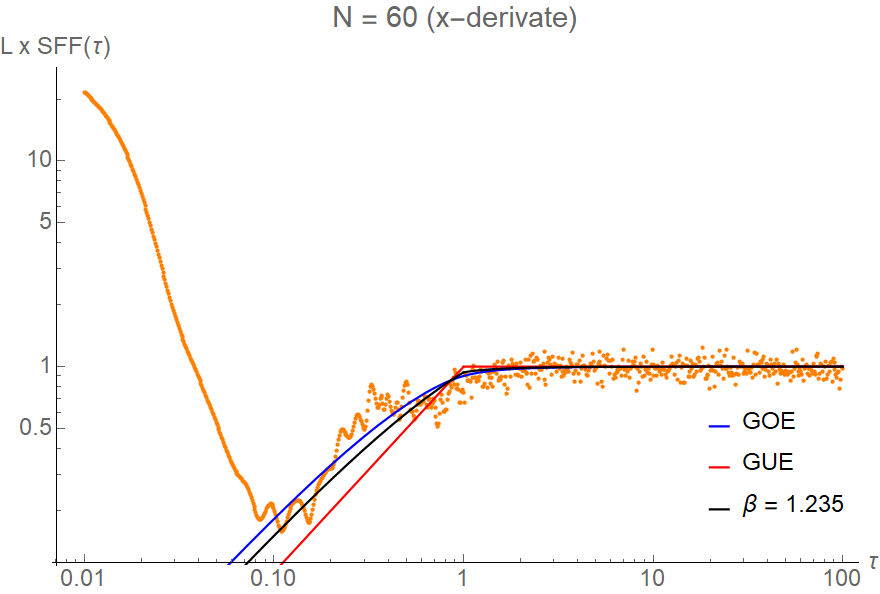}
\includegraphics[width=0.36\textwidth]{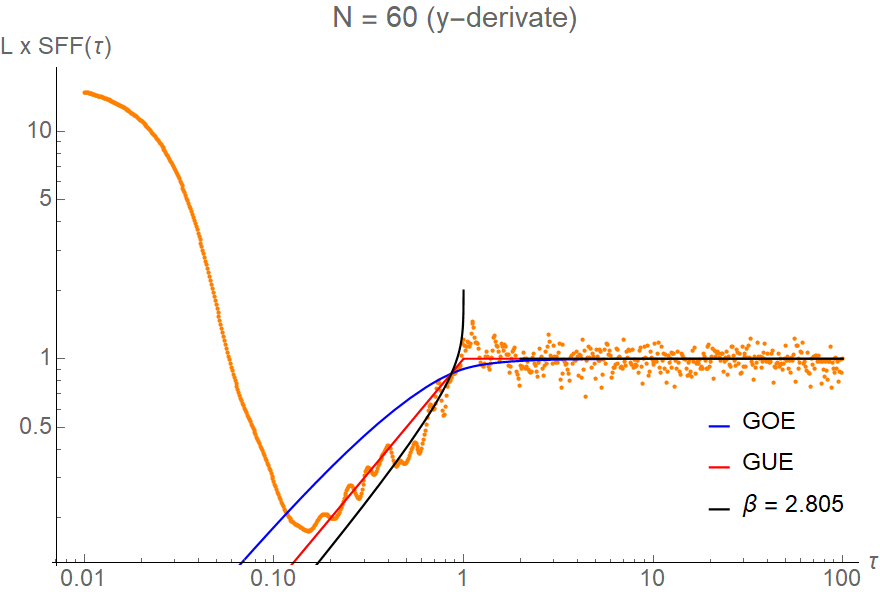} 

\includegraphics[width=0.36\textwidth]{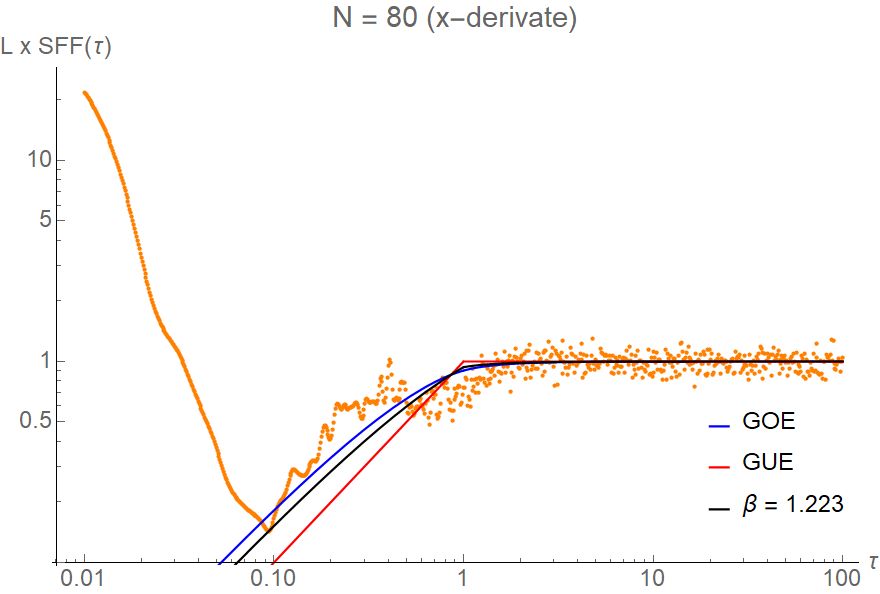}
\includegraphics[width=0.36\textwidth]{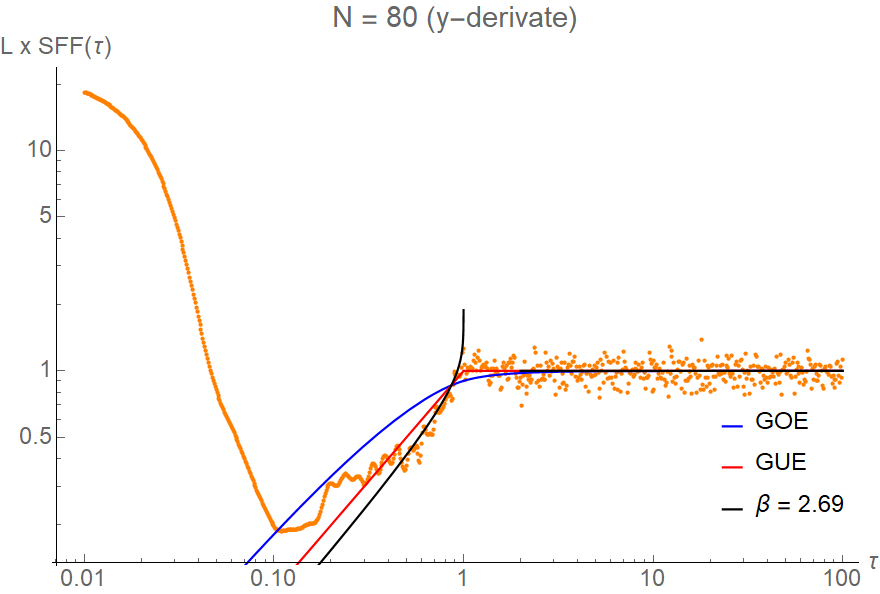} 

\includegraphics[width=0.36\textwidth]{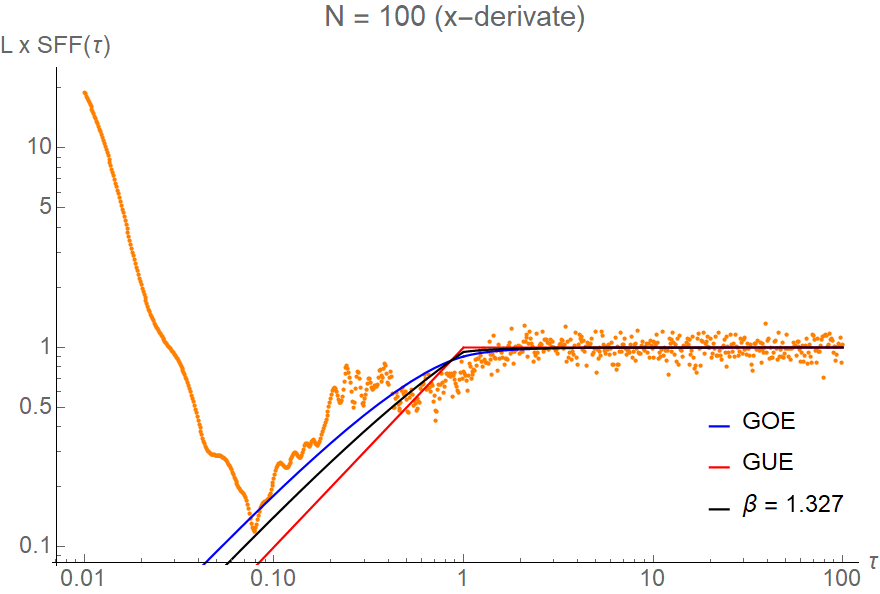}
\includegraphics[width=0.36\textwidth]{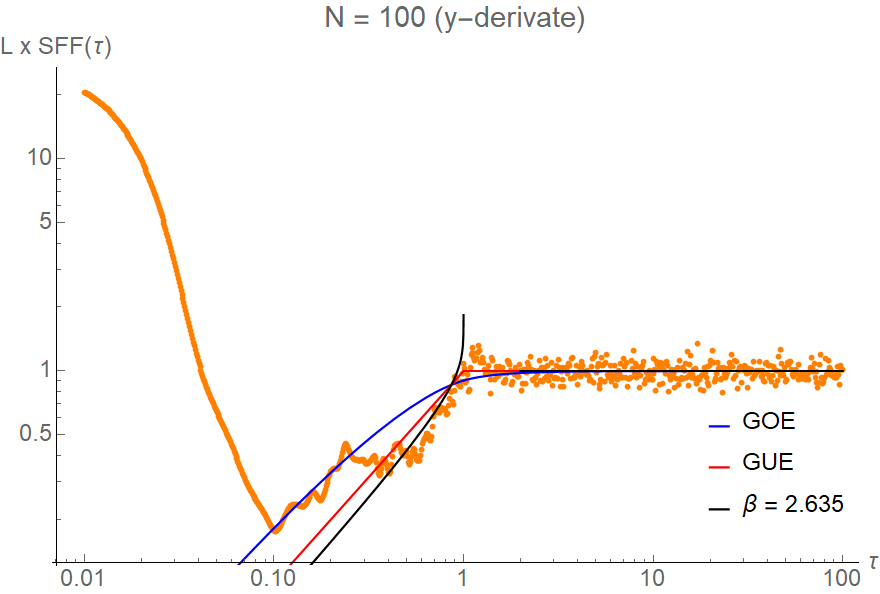} 

    \caption{The full AFF on a logarithmic scale for $N = 40$, 50, 60, 80, and 100. The left and right columns show the distributions associated with the $x$- and $y$-derivatives, respectively.}
    \label{fig:allsfflog}
\end{figure}

In figure \ref{fig:allsfflog} we plot the full AFF, \textit{i.e.} $R_2(\tau)$ defined from eq.~\eqref{eq:sff} without the subtraction of eq.~\eqref{eq:sffconn}, for all the states in our sample, on a log-log scale. This shows most clearly the decline-ramp-plateau structure of RMT.

\end{document}